\documentclass[9pt,twocolumn,twoside]{oljnl}
\journal{opticajournal} 
\setboolean{shortarticle}{true}
\usepackage{xcolor}
\title{One-sided composite cavity on an optical nanocapillary fiber}
\author{Srinu Gadde, Jelba John, and Ramachandrarao Yalla*}
\affil{School of Physics, University of Hyderabad, Hyderabad, Telangana 500046, India}
\affil[*]{rrysp@uohyd.ac.in}

\begin{abstract}
We numerically report a one-sided cavity on an optical nanocapillary fiber (NCF) using a composite cavity. The composite cavity is formed by combining an optical NCF and an asymmetric defect mode grating. We design the cavity to realize the maximum channeling efficiency of up to $\sim 80\%$ into one-sided NCF-guided modes while operating from under- to critical- and over-coupling regimes.  For the maximum channeling efficiency case, we found the best quality factor, finesse, and one-pass loss of the cavity are $19354$, $240$, and $1.3\%$, respectively. The present platform may open a novel route for designing fiber-based deterministic single-photon sources in quantum technologies. 
\end{abstract}
\setboolean{displaycopyright}{false}
\begin{document}
\maketitle
\section{Introduction}
In quantum internet and quantum information technologies, single photons are proven to be an ideal choice for efficient information carriers between the nodes of quantum network \cite{kimble2008quantum,flamini2018photonic,sunami2025scalable}. However, generating single photons alone is ineffective without efficient collection and controlled manipulation. In recent quantum computation techniques, high photon collection efficiency is required for the generation of high-rate entanglement \cite{sunami2025scalable}. In this direction, cavity quantum electrodynamics offer high-field confinement within the cavity by circulating the light \cite{kimble1998strong,thompson2013coupling,reiserer2015cavity,gallego2018strong}. The spontaneous emission (SE) of a single quantum emitter (SQE) is modified by placing it within the cavity structure. Various cavity schemes have been proposed and experimentally demonstrated to improve single photon collection efficiency. Examples are microcavities \cite{solomon2001single, vahala2003optical}, low mode area nanophotonic waveguides \cite{thompson2013coherence,hood2016atom,gonzalez2024light}, and photonic crystal (PhC) cavity structures \cite{englund2005controlling,hung2013trapped,goban2014atom,gonzalez2015subwavelength}. 

Optical nanofibers (ONFs) are emerging as versatile candidates for efficient information transmission, as they can be seamlessly integrated with single-mode fibers (SMFs) and also offer high evanescent fields in the sub-wavelength region and enable robust interactions with the surrounding medium \cite{vetsch2010optical,kato2015strong,nayak2018nanofiber}. To further enhance the collection efficiency of single photons from an SQE and subsequent coupling with the SMFs, different cavity structures are proposed and demonstrated experimentally in the case of ONFs. Examples are fiber Bragg grating (FBG)/PhC cavities \cite{kato2015strong,nayak2018nanofiber,qing2019simple,nayak2019real,samutpraphoot2020strong, ruddell2020ultra,sahu2022optimization,horikawa2024high} and composite PhC cavities (CPCC) \cite{yalla2014cavity,keloth2017fabrication}. In the former case, the cavity is formed on the ONF surface, while in the latter case, the cavity is realized by integrating an external nano-grating with the tapered region of ONF. There are two advantages to the CPCC technique. One is the precise positioning of an SQE at the cavity anti-node position \cite{yalla2014cavity} such that its SE is enhanced, and the other is tunability of cavity mode resonance wavelength \cite{schell2015highly,yalla2020design}. However, the maximum channeling efficiency ($\eta$) through the ONF-guided modes is limited to only 32\% - 43\% into either side of ONF-guided modes \cite{nayak2019real,sahu2022optimization,yalla2014cavity,li2018tailoring} and total SE of the SQE is channeled equally into two sides of ONF-guided modes. In all these cases, the SQE is placed on the ONF surface, which limits the interaction.  

To further improve the $\eta$-value, a new type of hollow core fiber termed capillary fiber was proposed and experimentally demonstrated. However, the maximum $\eta$-value is limited to 18\% \cite{faez2014coherent}. The lower $\eta$-value is attributed to the thicker cladding diameter relative to the core. For further enhancement of $\eta$-value, our group proposed and demonstrated a new type of capillary fiber, an optical nanocapillary fiber (NCF), featuring a core with sub-wavelength inner and outer diameters. When radially polarized SQE is placed inside the NCF, the maximum $\eta$-value reaches up to 52\% \cite{elaganuru2024highly}. We also reported that $\eta$-value up to 87\% using a composite photonic crystal symmetric cavity structure, which was realized by combining a defect mode symmetric nano-grating with the NCF \cite{gadde2025cavity}. In all these techniques, the total SE of the SQE is channeled equally into two sides of NCF-guided modes. However, several quantum information protocols have been proposed that utilize one-sided cavity schemes such as quantum phase switches with single atom and single photon transistor \cite{reiserer2015cavity,tiecke2014nanophotonic,tiarks2014single}. It has been experimentally demonstrated that the maximum $\eta$-value of 65\% is realized using a composite PhC asymmetric cavity on the ONF \cite{yalla2022one}. This technique allows the total SE to be channeled into one-sided ONF-guided modes.  

In this letter, we numerically report a one-sided cavity on the NCF using a composite cavity. The composite cavity is formed by combining an optical NCF and an asymmetric defect mode grating. We design the cavity to realize the maximum $\eta$-value of up to $\sim 80\%$ into one-sided NCF-guided modes while operating from under- to critical- and over-coupling regimes. For the maximum channeling efficiency case, we found the best quality factor, finesse, and one-pass loss of the cavity are $19354$, $240$, and $1.3\%$, respectively. The present platform may open a novel route for designing fiber-based deterministic single-photon sources in quantum technologies. 

\section{Methodology}
\begin{figure}[ht]
	\centering
	\includegraphics[width= 8 cm]{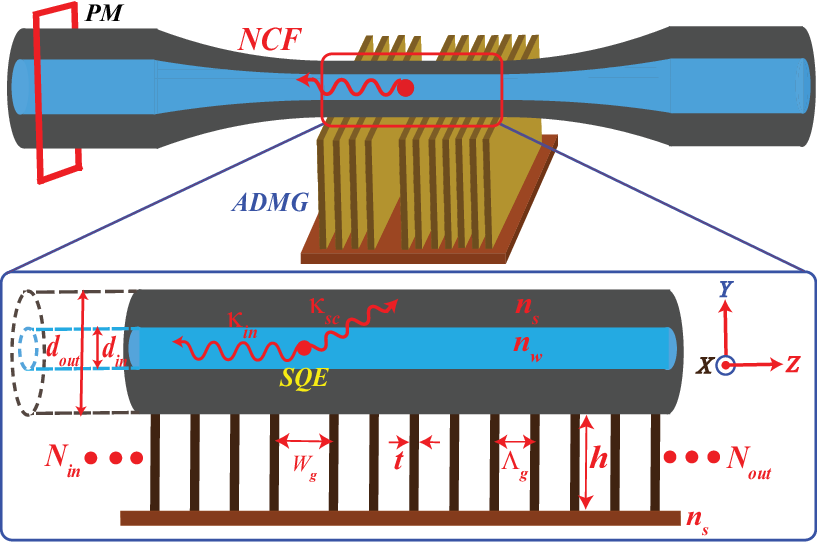}
	\caption{A conceptual diagram of a one-sided composite cavity. The cavity is realized by combining an asymmetric defect mode grating (ADMG) with an optical nanocapillary fiber (NCF), and $y$-polarized single quantum emitter (SQE) placed at the cavity anti-node position. The inset shows the side view of the cavity with the designed parameters as mentioned in the text. $\kappa_{in}$ and $\kappa_{sc}$ are the input coupling rate and scattering rate (intra-cavity loss rate) of the cavity, respectively.} \label{fig1}
\end{figure}

We utilize Ansys 3D FDTD software to carry out all simulations numerically. We design a composite cavity in such a way as to channel the total SE from the SQE into one-sided NCF-guided modes. For this purpose, a one-sided PhC cavity is designed by combining an optical NCF with an asymmetric defect-mode grating (ADMG), and a $y$-polarized SQE is placed at the cavity anti-node position, as shown in Fig. \ref{fig1}. The designed parameters of the one-sided cavity are sub-wavelength inner ($d_{in}$) and outer ($d_{out}$) diameters of cylinders, duty cycle ($\alpha$), slat thickness ($t$= $\alpha \Lambda_g$), slat height ($h$), grating period ($\Lambda_g)$, width of the defect ($w_g$= $1.5\Lambda_g $), slat number at the output ($N_{out}$), and slat number at the input ($N_{in}$) side of the ADMG are shown in the inset of Fig. \ref{fig1}. Note that we consider refractive indices of silica ($n_{s}$) and water ($n_{w}$) media for the inner and outer cylinders of the NCF. As per Ref. \cite{gadde2025cavity}, the optimum parameters for a two-sided cavity are $\Lambda_g$= $244$ nm, $w_g$= $366$ nm, $\alpha$= $15\%$, total slat number $N$ (= $N_{out}$ + $N_{in}$) 400, $t$= $36.6$ nm, $h$= 2 $\mu m$, and $d_{in}$ ($d_{out}$)= 125 (515) nm. The power coupled to the NCF-guided modes ($P_{g}$) is determined using a power monitor (PM) from the SQE to the left side of a cavity as shown in Fig. \ref{fig1}. We find the $\eta$-value as $\eta$= ${P_g}/{P_c}$ and the Purcell factor ($F_p$) as $F_p$= ${P_c}/{P_0}$. $P_c$ and $P_0$ are the power emitted by the SQE in presence of a cavity and free space, respectively. The whole design of the composite system, enclosed by a three-dimensional simulation region of $3$$\times$4.5$\times$250 $\mu m^3$ within the perfectly matched layers as boundary walls.
The essential point is the unequal slat number ($N_{in}$$\neq$$N_{out}$) on either side of the ADMG to realize the unidirectional channeling of the total SE of the SQE. For this purpose, we find the optimum $N_{out}$, $N_{in}$-values by simulating $\eta$ and $F_P$-values while keeping scattering losses as low as possible while achieving maximum $\eta$-value through one-sided NCF-guided modes. In the first case, we optimize the $N_{out}$ and $N_{in}$-values while monitoring $\eta$ and $F_P$-values. Subsequently, we examine the reflection spectra from the under- to critical- and over-coupling regimes. Finally, we analyze the on-resonance reflectivity ($R_0$) and the corresponding total cavity line widths ($\kappa$) for $y$-, $x$-polarized input mode sources.

\section{Results and Discussions}
\begin{figure}[ht]
	\centering
	\includegraphics[width=\linewidth]{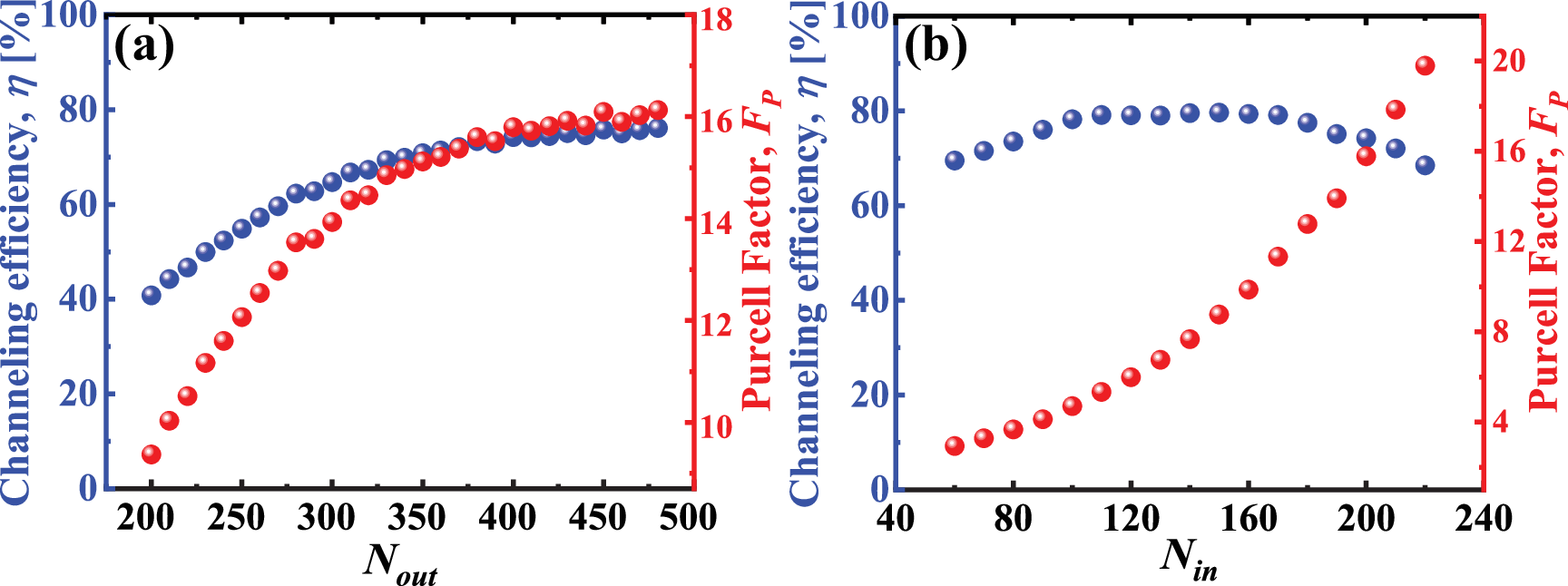}
	\caption{Depict the dependency of channeling efficiency ($\eta$, blue spheres), Purcell factor ($F_p$, red spheres) on (a) output ($N_{out}$) slat number while input ($N_{in}$) slat number is fixed at 200 and (b) input ($N_{in}$) slat number while output  ($N_{out}$) slat number is fixed at 400.} \label{fig2}
\end{figure}

\begin{figure}[ht]
	\centering
	\includegraphics[width= 8 cm]{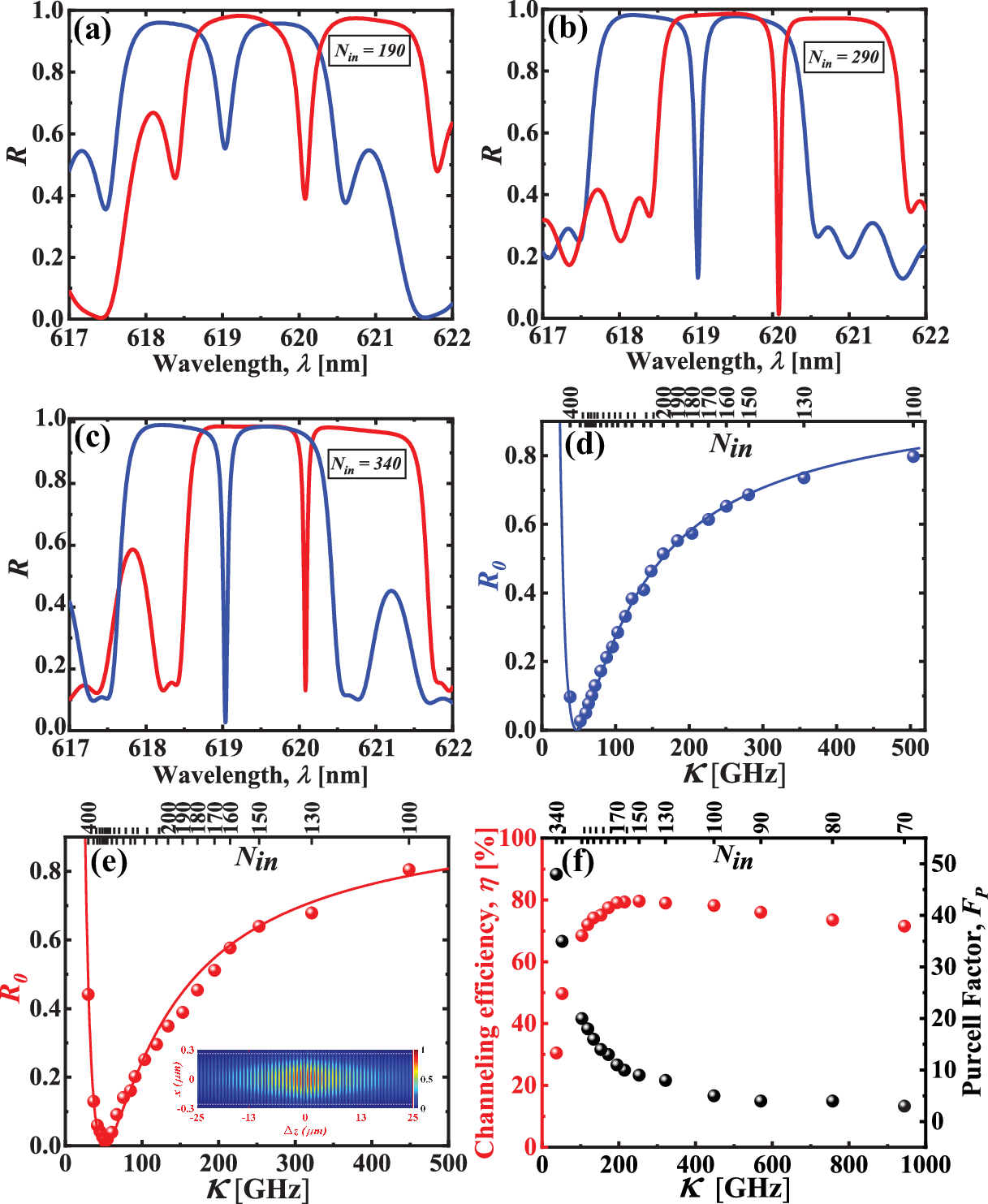}
	\caption{(a), (b) and (c) are the typical cavity reflection spectra for $x$ (blue trace)- and $y$ (red trace)-polarized input mode sources for the input slat number ($N_{in}$) at 190 (over), 290 (critical), and 340 (under) for the fixed output slat number ($N_{out}$) at 400, respectively. (d) and (e) are the on-resonance cavity's reflectivity ($R_0$) for $x$ (blue spheres) and $y$ (red spheres)-polarized input mode sources, respectively. The top $x$-axes represent the corresponding $N_{in}$-values. The solid blue (red) line fits the data for $x$ ($y$)-pol. The insets show the normalized electric field intensity distribution at the anti-node ($\Delta z$= 0 nm) of the cavity, and white dotted lines represent the NCF surfaces. (f) depicts the variation of channeling efficiency ($\eta$, red spheres), purcell factor ($F_p$, black spheres) with the $\kappa$-values and the corresponding $N_{in}$-values are shown in the top $x$-axis.}\label{fig3}
\end{figure}

Figures \ref{fig2} (a) and (b) depict the dependency of $\eta$-value (blue spheres) and $F_P$-value (red spheres) on $N_{out}$ and $N_{in}$-values, respectively. The horizontal axes correspond to $N_{out}$ and $N_{in}$-values, while the left and right vertical axes correspond to $\eta$- and $F_P$-values, respectively. In Fig. \ref{fig2} (a), we fixed the $N_{in}$-value at 200, and strength of the cavity's right side mirror is increased by increasing the $N_{out}$-value from 200 to 480 in steps of 10 while monitoring the $\eta$, $F_P$-values. One can see that $\eta$-value and $F_P$-value gradually increase while increasing $N_{out}$-value.  Note that no significant increase is observed in either parameter for $N_{out}$$\geq400$. Therefore, we fix the $N_{out}$-value at 400. In Fig. \ref{fig2} (b), we increase the strength of the cavity's left side mirror by increasing the $N_{in}$-value from 60 to 220 in steps of 10 while monitoring the resulting $\eta$, $F_P$-values. One can readily see that the $\eta$-value increases gradually up to 110 and remains relatively constant between 110 to 170, and then begins to decrease for $N_{in}$$\geq170$. This is attributed to reflecting the NCF-coupled spontaneous emission rate of an SQE coming out of the cavity. However, $F_P$-value continues to increase as expected, indicating a steady enhancement in the spontaneous emission rate of SQE. This suggests that, despite the increased emission rate, the $\eta$-value into the NCF-guided modes is decreasing. Hence, the maximum $\eta$-value of $80\%$ through one-sided NCF-guided modes is obtained at the optimum $N_{in}$= $150$ and $N_{out} \geq 400$.

First, we analytically examine the coupling characteristics of the designed cavity. The reflectivity ($R_0$) under the on-resonance condition is $R_0$= $\bigg|\dfrac{\kappa_{in}-\kappa_{sc}}{\kappa} \bigg|^{2}$= $\bigg|1-{\dfrac{2 \kappa_{sc}}{\kappa}} \bigg|^{2} $\cite{yalla2022one}. $\kappa$ (= $\kappa_{in}+ \kappa_{sc}$), $\kappa_{in}$, and $\kappa_{sc}$ are the total decay rate, input coupling rate, scattering rate (intra-cavity loss rate) of the cavity, respectively, as illustrated in Fig. \ref{fig1}. It is clear that while decreasing $\kappa$-value, $R_0$-value exhibits distinct behaviors across different coupling regimes: it decreases in the over-coupling regime (${\kappa}/{2}>\kappa_{sc}$), where scattering losses are less dominant; reaches to zero at critical-coupling regime (${\kappa}/{2}$= $\kappa_{sc}$), where both intra-cavity losses and total cavity decay rates are balanced; and increases in the under-coupling regime (${\kappa}/{2}<\kappa_{sc}$), where intra-cavity losses are dominant.

To investigate the cavity's characteristics, we simulate its reflection spectra for different $N_{in}$-values of 190, 290, and 340 at the fixed $N_{out}$= 400 for $x$ (blue trace) and $y$ (red trace) polarized input mode source and the corresponding results are shown in the Figs. \ref{fig3} (a), (b), and (c), respectively. These cases correspond to the over-, critical, and under-coupling regimes of the cavity. The horizontal and vertical axes correspond to the wavelength (\textit{$\lambda$}) and reflectivity ($R$), respectively. One can readily see a strong photonic reflection band along with a dip at the center of a band around the designed cavity mode resonance wavelength ($\lambda_{0}$) of 620 (619) nm for $y$ ($x$)-polarized input mode source. We determine the $\lambda_{0}$-value and corresponding cavity line widths ($\Delta \lambda$) and on-resonance reflectivity ($R_0$) by fitting a dip in reflection with a Lorentzian. It is clear that $\Delta \lambda$-values for $y$-pol. are narrower compared to $x$-pol. This is attributed to the cavity's $y$-pol. mode experiences a larger refractive index modulation due to the DMG along the $y$-direction. Consequently, $Q$-values for $y$-pol. are higher compared to $x$-pol., owing to the higher reflectivity induced by the DMG along the $y$-direction of the cavity \cite{yalla2014cavity}. Additionally, for the case of $N_{in}$= 340, a slight asymmetry is observed wherein the $R_0$-value for $y$-pol. begins to increase. This occurs because the cavity mirrors inhibit the entry of input light, and hence most of the input light is reflected back through NCF-guided modes. It should be mentioned that $\lambda_{0}$-value can be tuned up to $\pm 10$ nm \cite{yalla2020design}. Additionally, we simulate the reflection spectra for $N_{in}$-values from 100 to 400 by keeping $N_{out}$-value fixed at 400 and then determine the corresponding $R_0$, $\kappa$-values. The results are shown in Figs. \ref{fig3} (d) and (e) for $x$-pol and $y$-pol, respectively. It depicts the variation of $R_0$-value as a function of $\kappa$-values. The horizontal axis represents the $\kappa$-value (bottom) and the $N_{in}$-value (top), while the vertical axis represents the $R_0$-value. The blue (red) spheres indicate $x$ ($y$)-polarized input modes. One can also observe that, as $N_{in}$-value increases, the $\kappa$-value decreases, and simultaneously, $R_0$-value first decreases and reaches 0 and again increases. It reflects the cavity's behavior transition from over- to critical- and under-coupling regimes. 

To estimate the scattering limited decay rate ($\kappa_{sc}$= $\kappa/2$, when $R_0 \approx$ 0), we fit the equation for $R_0$-value with $\kappa_{sc}$-value as a free parameter. The corresponding results are shown in the Tab. \ref{Table1} for the $y$-pol. To extract the effective length of the cavity ($l_{eff}$), we simulated the intensity profile of the cavity mode and showed in the inset of Fig. \ref{fig3} (e). It shows the normalized electric field intensity distribution in the $xz$-plane at $y$= 0 for $y$-polarized SQE placed at the cavity anti-node ($\Delta z$= 0 nm). White dotted lines represent the NCF surfaces. We found the $l_{eff}$-value to be 25 $\mu m$. Using $\kappa_{sc}$ and $l_{eff}$-values, we determine the cavity's performance parameters such as scattering-limited quality factor ($Q_{sc}$), finesse ($\mathcal{F}_{sc}$), and one-pass power loss ($\mathcal{L}$) \cite{yalla2022one}. Table \ref{Table1} shows the comparison of the cavity's performance parameters with the other one-sided cavity works in the case of ONF for the maximum $\eta$-value case. It is clear that the $\eta$-value is on par with other structures \cite{sahu2024slot}. One can also observe that the $\kappa_{sc}$-value is lower compared to other structures \cite{yalla2022one, sahu2024slot} while the scattering limited $Q_{sc}$ and $\mathcal{F}_{sc}$-values are higher than other structures \cite{yalla2022one, sahu2024slot} whereas the one-pass $\mathcal{L}$-value is lower than the quoted values in Ref. \cite{yalla2022one}. The maximum scattering-limited $Q_{sc}$-value and minimum one-pass $\mathcal{L}$-values are estimated to be around 19354$\pm$178 and 1.3$\pm$0.01\%, respectively. Note that we also determined the $\kappa_{sc}$-value for the $x$-pol and found that it is lower compared to the $y$-pol. This is due to lower scattering loss induced by the ADMG along the $x$-direction compared to $y$-direction.

\begin{table}[htbp]
		\centering
		\caption{Comparison with different one-sided cavity schemes} \label{Table1}
		\begin{tabular}{|p{0.2cm}|p{0.3cm}|p{0.3cm}|p{0.4cm}|p{0.6cm}|p{0.55cm}|} 
			\hline
			\multicolumn{1}{|p{0.5cm}|}{Refs.$\downarrow$}
			&\multicolumn{1}{|p{0.8cm}|}{$\eta$[\%]}
			&\multicolumn{1}{|p{1.2cm}|}{$\kappa_{sc}$[GHz]}
			&\multicolumn{1}{|p{0.6cm}|}{$Q_{sc}$} 
			&\multicolumn{1}{|p{0.2cm}|}{$\mathcal{F}_{sc}$}
			&\multicolumn{1}{|p{0.2cm}|}{$\mathcal{L} [\%]$} \\
			\hline
			\multicolumn{1}{|p{0.5cm}|}{\cite{yalla2022one}}
			&65&54&8680&124&2.5\\
			\hline
			\multicolumn{1}{|p{0.5cm}|}{\cite{sahu2024slot}}
			&86&59&6388&253&1.2\\ 
			\hline
			\multicolumn{1}{|p{0.8cm}|}{Present} 
			&80&25&19354&240&1.3\\
			\hline
		\end{tabular}
	\end{table}
Next, we examine the variation of $\eta$/$F_P$-values with the corresponding $\kappa$-values in three coupling regimes as shown in Fig. \ref{fig3} (f). The horizontal axis bottom (top) represents the $\kappa$ ($N_{in}$)-values, while the vertical axis left (right) represents the $\eta$ ($F_P$)-values. One can readily see that $\eta$-value more than 70\% is realized in the over-coupling regime ($\kappa$ $\geq$ 100 GHz) of the cavity. It is also clear that, $F_P$-value increases monotonically while increasing the $N_{in}$-value. However, $\eta$-value initially increases upto $N_{in}$= 150 and again decreases. The $\kappa$-values are in the range of GHz, the present cavity can be effectively operated in the Purcell regime of cavity-QED \cite{yalla2014cavity, keloth2017fabrication}. Therefore, the $F_{P}$ can be approximately equal to cooperativity ($C$). Using the $F_{P}$ (15.8) and $\kappa$ (253 GHz)-values at the optimum condition, we have estimated the atom-cavity coupling rate, $2g_{0} \approx$ 70 GHz by assuming the spontaneous emission decay rate ($\gamma$) of 1.2 GHz in the case of NV centers in nanodiamond \cite{zhao2013observation}. It should be mentioned that suitable SQEs are high quantum efficiency semiconductor quantum dots and NV/SiV centers in nanodiamonds.

Regarding the experimental realization of the designed cavity, one is the fabrication of a tapered optical NCF with the prerequisite inner and outer diameters. This can be achieved by tapering the commercially available capillary fiber using heat and a pulling technique \cite{bashaiah2024fabrication}. The other is the fabrication of ADMG with the required input and output slat number about the defect center, which can be realized with the electron beam lithography technique \cite{yalla2022one}. The critical challenge is the precise positioning of an SQE at the cavity's anti-node position. This can be addressed by injecting the water with dissolved quantum dots as SQE by using either a capillary force or a peristaltic pump technique while fine-tuning the flow parameters \cite{white2006liquid}. The cavity formation is achieved by integrating the ADMG onto the waist region of a tapered NCF. It is worth noting that a key advantage in the composite cavity case is the flexibility in aligning the ADMG relative to the NCF to maximize the light-matter interaction \cite{yalla2014cavity,schell2015highly,yalla2022one}. It was already demonstrated that the experimental realization of a one-sided composite cavity in the case of the ONF \cite{yalla2022one}. It is worth mentioning that the placement of an SQE at any one of the anti-nodes within a uniform region, as shown in the inset of Fig. \ref{fig3} (e), leads to the same spontaneous emission enhancement. This allows experimental feasibility in the positioning of SQE at the cavity anti-node.

The present $\eta$-value can be enhanced in two different ways. One is replacing the water medium with any higher refractive index liquid \cite{faez2014coherent}. However, experimentally, the flow of such a high-refractive-index liquid with dissolved quantum dots as SQE through the NCF may be challenging. The other is replacing silica material (outer diameter of NCF) with any higher refractive index materials such as diamond, gallium phosphate, or silicon nitride \cite{resmi2024highly}. Hence, the present cavity design can be effectively implemented for efficient unidirectional channeling of total spontaneous emission through NCF-guided modes and also works as a deterministic single-photon source. It is worth noting that unidirectional channeling of single photons can also be achieved through chiral light-matter interactions; however, this approach only allows control of the photon flux in one direction.

\section*{Conclusion}
In summary, we numerically reported a one-sided cavity on an optical nanocapillary fiber (NCF) using a composite technique. The cavity was formed by combining the optical NCF and an asymmetric defect mode grating. We realized the maximum channeling efficiency of up to $\sim 80\%$ into one-sided NCF-guided modes while operating from under- to critical- and over-coupling regimes. For the maximum channeling efficiency case, we found the best quality factor, finesse, and one-pass loss of the cavity are $19354$, $240$, and $1.3\%$, respectively. The present platform may open a novel route for designing fiber-based deterministic single-photon sources in quantum technologies. 
\begin{backmatter}
\bmsection{Acknowledgments} RRY acknowledges financial support from the Scheme for Transformational and Advanced Research in Sciences (STARS) grant from the Indian Institute of Science (IISc), Ministry of Human Resource Development (MHRD) (File No. STARS/APR2019/PS/271/FS) and Institute of Eminence (IoE) grant at the University of Hyderabad, Ministry of Education (MoE) (File No. UoH-IoE-RC2-21-019).
	
\bmsection{Disclosures} The authors declare no conflicts of interest.
	
\bmsection{Data availability}
Data underlying the results presented in this paper are not publicly available at this time but may be obtained from the authors upon reasonable request.
\end{backmatter}

\bibliography{bibliography}
\bibliographyfullrefs{bibliography}
\end{document}